\documentclass[
    ,final            
    ,sort&compress                 
  ]
  {aipproc}

\usepackage{hyperref}

\layoutstyle{8x11single}

\begin{document}

\title{B-Tagging at CDF and D\O \\ Lessons for LHC}


\author{Thomas Wright\thanks{Speaker, on behalf of the CDF 
	and D\O~collaborations}}{
  address={University of Michigan \\ Ann Arbor, MI  48109}
}

\begin{abstract}

The identification of jets resulting from the fragmentation and
hadronization of $b$ quarks is an important part of high-$p_T$
collider physics.  The methods used by the CDF and D\O\ collaborations
to perform this identification are described, including the calibration of
the efficiencies and fake rates.  Some thoughts on the application of
these methods in the LHC environment are also presented.

\end{abstract}

\maketitle


\section{Introduction}

The identification or ``$b$-tagging'' of jets resulting from the
fragmentation and hadronization of $b$ quarks is an important part of
high-$p_T$ collider physics.  Some examples include the study of top
quark production or searches for low-mass Higgs bosons in the dominant
$b\bar{b}$ decay mode.

Jets containing $b$ hadrons have distinctive properties that are
useful in distinguishing them from other types of jets.  One is a
relatively high rate of lepton production from semileptonic decays.
Because the fragmentation is hard and the $b$ hadrons retain about
70\% of the original $b$ quark momentum, the leptons will generally
have high $p_T$ relative to the jet $p_T$, which makes them easier to
identify and separate from lepton sources in generic jets such as
decays in flight of $\pi$'s or $K$'s.  The large mass of $b$ hadrons
also helps, as leptons from $B$ decays will have $\sim 1$ GeV/$c$ of
$p_T$ relative to the jet axis, while leptons and fakes in generic
jets tend to be more closely aligned with the jet.

A second property of $b$ hadrons useful for tagging is their long
lifetime, with $c\tau \sim 450$ $\mu$m.  A $b$ hadron with $p_T = 50$
GeV/$c$ will then fly on average almost half a centimeter before
decaying.  The large mass of the hadron produces enough opening angle
that the daughter particles can have sizable impact parameters with
respect to the $b$ hadron point of origin.  These particles also have
high $p_T$, which reduces the effects of multiple scattering and
allows these impact parameters to be measured with good resolution.

The CDF and D\O\ detectors are described in
Refs.~\citep{Acosta:2004yw,Abazov:2005pn}.  Both feature a
high-efficiency central tracker with good momentum resolution,
surrounding a silicon strip detector for precise position
measurements.  Impact parameter resolutions are typically 40-60 $\mu$m
depending on the track $p_T$, including a 30 $\mu$m contribution from
the beam width.  

\section{Soft Lepton Tagging}

\begin{figure}
  \includegraphics[width=0.6\textwidth]{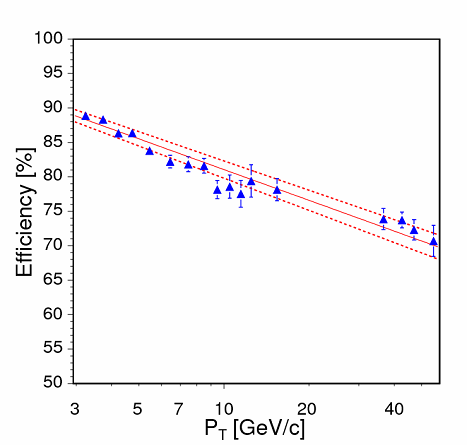}
  \caption{Efficiency of the CDF soft muon identification as measured
  from $J/\psi$ and $Z$ decays, as a function of the muon $p_T$.}
  \label{f:cdf-slt}
\end{figure}

As previously mentioned, the presence of leptons is a good signature
of the presence of $b$ hadrons in a jet.  The key is to define an
identification algorithm that maintains good performance even in the
busy environment around the center of the jet.  That means that
quantities which are typically used for high-$p_T$ lepton selection,
such as calorimeter energy deposition consistent with electrons or
muons, cannot generally be used because of the presence of other
particles nearby.  For this reason muons are somewhat preferred, as
most of the identification occurs outside of the calorimeter after the
surrounding particles have ranged out.  Both CDF and D\O\ have
published results using a soft muon tagger
\citep{Acosta:2005zd,Abazov:2004bv}.  Figure~\ref{f:cdf-slt} shows the
per-muon efficiency of the CDF soft muon tagger as a function of the
muon $p_T$, measured using the second legs of $J/\psi$ and $Z$
decays.  When the branching ratios and $p_T$ spectra are included the
net tagging efficiency per $b$ jet is in the 10\% range.  Fake rates
measured from generic jet samples are about 0.5\% per muon candidate.

\section{Lifetime Tagging}

\begin{figure}
  \includegraphics[width=0.6\textwidth]{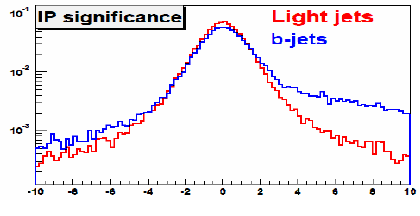}
  \caption{Impact parameter significance in light-flavor and $b$ jets
  at D\O.}
  \label{f:d0-imp-sig}
\end{figure}

Tagging algorithms which exploit the long lifetime of $b$ hadrons have
an advantage over the lepton taggers in that they are more inclusive
and not limited by semileptonic branching ratios.  The basic
ingredient of a lifetime tagger is measuring the impact parameters of
the tracks within a jet.  Figure~\ref{f:d0-imp-sig} shows the
distributions of impact parameter significance (impact parameter
divided by its estimated uncertainty) for light-flavor and $b$ jets at
D\O.  The impact parameters are signed such that tracks which cross
the jet axis behind the primary vertex relative to the jet direction
are negative.  $b$-jets show a clear excess of tracks with significant
positive displacement.

\begin{figure}
  \includegraphics[width=0.6\textwidth]{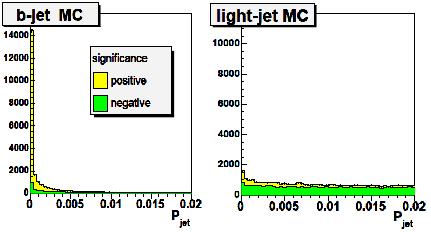}
  \caption{Probability distributions for $b$ jets (left) and light
  jets (right), using tracks with positive or negative impact
  parameters.}
  \label{f:d0-jet-prob}
\end{figure}

One way to use the impact parameter information is to form the joint
probability for all tracks in the jet to have originated from the
event primary vertex.  This is done using distributions similar to the
``light jets'' curve in Figure~\ref{f:d0-imp-sig} as probability
density functions.  The tracks with high impact parameter significance
which occur in $b$-jets will cause this joint probability to peak at
low values, as shown in Figure~\ref{f:d0-jet-prob}.  The probability
cut can be tuned to obtain a desired purity or efficiency.  More
information on the CDF and D\O\ implementations of this algorithm can
be found in Refs.~\citep{Abulencia:2006kv,Greder:2004sa}.

Another way to use the high impact parameter tracks is to require that
they be consistent with originating from a secondary displaced
vertex.  Typically, tracks above a $p_T$ threshold (1-2 GeV/$c$) and
with high impact parameter significance (2-3) are fit to a common
vertex.  A pruning algorithm removes tracks with high $\chi^2$
contribution until a set threshold is reached or there are no
tracks left.  Finally, a cut on the vertex displacement significance
is applied to obtain the desired fake rate.  Both experiments have
implementations of this
algorithm~\citep{Acosta:2004hw,Abulencia:2006in,Abazov:2005ey}.

\subsection{Efficiency Measurement}

\begin{figure}
  \includegraphics[width=0.6\textwidth]{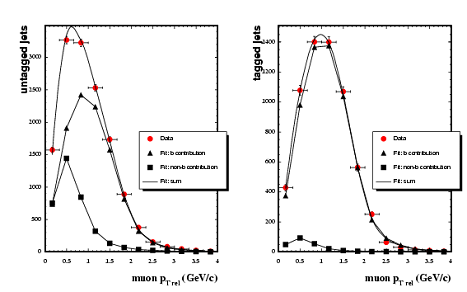}
  \caption{Distributions of muon $p_T$ relative to the jet axis at
  CDF, for untagged (left) and tagged (right) jets.}
  \label{f:cdf-mu-ptrel}
\end{figure}

Because of the difficulty of modeling the impact parameter
distributions in a detector simulation, it is important to measure the
efficiency of the taggers from real data.  Both experiments use
inclusive muon-triggered samples for this purpose.  The muon not only
enhances the $b$-fraction of the data, but its distinctive $p_T$
relative to the jet axis in $b$ jets allows that $b$-fraction to be
measured.  Figure~\ref{f:cdf-mu-ptrel} shows distributions of this
relative $p_T$ for untagged and tagged jets containing a muon in the CDF
data.  By fitting templates for the $b$ and non-$b$ components, the
numbers of untagged and tagged $b$ jets and hence the $b$-tagging
efficiency can be found.  D\O\ use a similar method, although instead
of fitting the $p_T$ distribution it is split into two bins and the
efficiency solved for algebraically.  CDF have a second method using
an electron sample, with the non-$b$ component inferred from the rate
of identified conversion pairs.

\begin{figure}
  \includegraphics[width=0.6\textwidth]{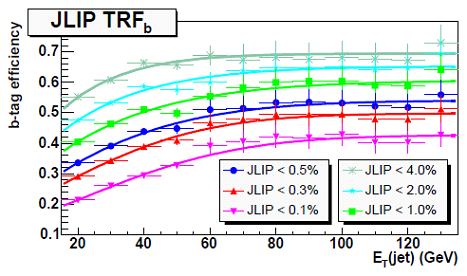}
  \caption{Parametrizations of the $b$-tagging efficiency for the D\O\
  jet probability tagger, derived from simulation and corrected to
  match the data.}
  \label{f:d0-jlip-eff}
\end{figure}

Because these jets containing leptons are not representative of
generic $b$-jets, the efficiencies measured in these samples cannot be
used directly.  Instead, samples of simulated events passing the same
cuts are generated, and a ratio of $b$-tagging efficiencies between
data and simulation, or ``scale factor'' is derived.  This scale
factor can then be used to correct the $b$-tagging efficiency in any
simulated sample to match the data.  In practice the scale factor is a
function of jet $E_T$ and $\eta$.  Figure~\ref{f:d0-jlip-eff} shows
some typical corrected efficiency parametrizations for the D\O\ jet
probability tagger.

\subsection{Fake Rate Measurement}

As with the efficiency, the tagger fake rates need to be determined
from the data.  Generally this is done using tracks with negative
impact parameters, which are unaffected by the presence of heavy
flavor.  The joint probability can be computed using negative tracks
instead of positive ones as shown in Figure~\ref{f:d0-jet-prob}, or
for the displaced vertex taggers the rate at which vertices are found
behind the primary vertex with respect to the jet direction can be
used as an estimate of the fake rate.

\begin{figure}
  \includegraphics[width=0.6\textwidth]{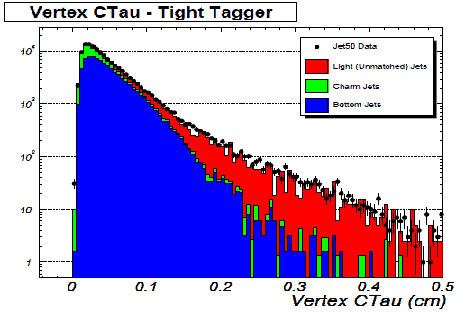}
  \caption{Net distribution of pseudo-$c\tau$ for the CDF displaced
	vertex tagger, derived by subtracting the inverted
	negative tagged distribution from the positive tagged one.}
  \label{f:cdf-ctau}
  \begin{picture}(10,10)
    \put(-290,70){\mbox{\rotatebox{90}{\bf \em Net POS-NEG tag count}}}
  \end{picture}
\end{figure}
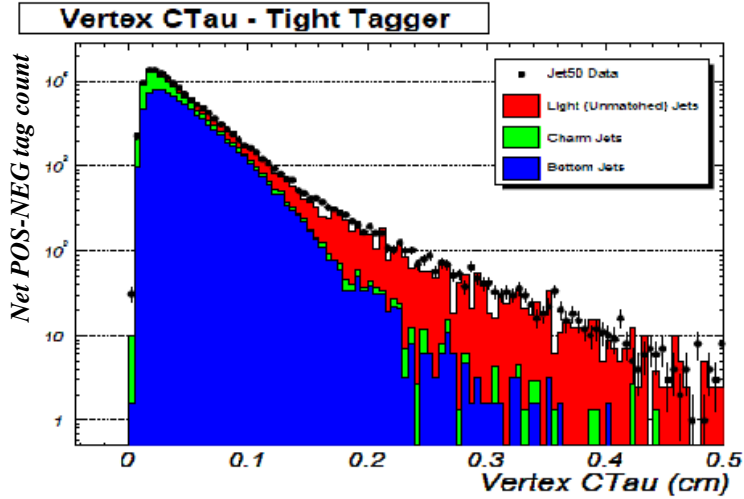

\begin{figure}
  \includegraphics[width=0.6\textwidth]{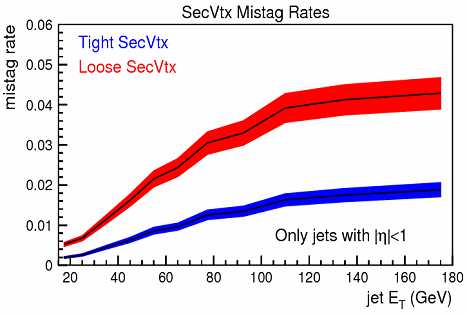}
  \caption{Postive fake tag rates for two tunes of the CDF displaced
  vertex tagger.}
  \label{f:cdf-mistag-et}
\end{figure}

These estimates account for fake tags due to misreconstructed tracks,
however they do not include tags from $K_S/\Lambda$ tracks surviving
the removal cuts or from interactions with the detector material, as
these will produce preferentially positive tracks and vertices.  These
effects can be estimated using the pseudo-$c\tau$
(defined as $L_{xy} \times M_{vtx}/P_{T,vtx}$) distribution of
vertices, after subtracting the symmetric fake component derived from
the negative tags, and fitting for the very long lifetime component as
shown in Figure~\ref{f:cdf-ctau} for the CDF displaced vertex tagger.
Based on this fit the negative tag rate must be scaled up by 30\% to
get the postive fake tag rate, shown in Figure~\ref{f:cdf-mistag-et}
as a function of jet $E_T$.

\section{Multivariate Taggers}

\begin{figure}
  \includegraphics[width=0.6\textwidth]{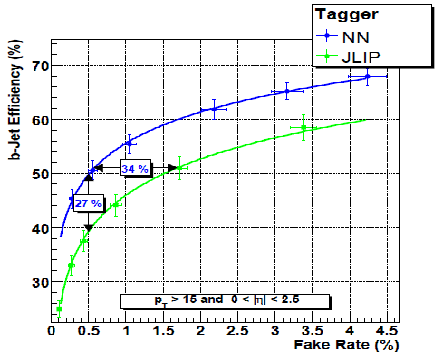}
  \caption{Efficiency and purity for the D\O\ multivariate tagger,
  compared to their joint probability algorithm.}
  \label{f:d0-nntag}
\end{figure}

Rather than simply cutting on the joint probability or the vertex
significance, a multivariate discriminant derived from these
quantities and other properties of the tags, such as the invariant
mass of the tracks in the vertex, the number of tracks in the vertex,
etc. can be constructed and used to select jets.  Both experiments
have released preliminary results~\citep{cdfwh,d04b} using such
taggers.  Including more information allows for either a more
efficient or higher-purity selection than is possible with the
single-variable taggers, as shown in Figure~\ref{f:d0-nntag} for the
D\O\ multivariate tagger.

\section{Summary}

CDF and D\O\ have developed high-performance and well-understood tools
for the identification of $b$ jets, and have used them to publish many
important physics results.  These tools will also be important at the
Large Hadron Collider when it starts running next year.  Much of the
experience gained at the Tevatron should be beneficial in
commissioning, such as how to choose track quality cuts, the
importance of a well-tuned simulation, and triggering strategies for
the data samples necessary for characterization of the tagger
performance.  The large leap in energy and luminosity will certainly
present additional challenges, but the physics potential guarantees
they will be solved.




\newpage 

\bibliographystyle{aipproc}   

\bibliography{thebibliography}



\end{document}